\documentclass[journal]{IEEEtran}
\usepackage[T1]{fontenc}
\usepackage[latin9]{inputenc}
\usepackage{float}
\usepackage{amsmath}
\usepackage{amsthm}
\usepackage{amssymb}
\usepackage{graphicx}
\usepackage{hyperref}
\usepackage{esint}
\usepackage{cleveref}
\makeatletter

\providecommand{\tabularnewline}{\\}
\floatstyle{ruled}
\newfloat{algorithm}{tbp}{loa}
\providecommand{\algorithmname}{Algorithm}
\floatname{algorithm}{\protect\algorithmname}

\theoremstyle{plain}
\newtheorem{thm}{\protect\theoremname}
\theoremstyle{remark}
\newtheorem{claim}[thm]{\protect\claimname}

\usepackage{cite}
\usepackage{psfrag}
\usepackage{url}

\usepackage{array}
\usepackage{epsfig}
\usepackage{multicol}
\usepackage{times}
\IEEEoverridecommandlockouts

\providecommand{\claimname}{Claim}
\providecommand{\theoremname}{Theorem}

\providecommand{\claimname}{Claim}
\providecommand{\theoremname}{Theorem}

\providecommand{\claimname}{Claim}
\providecommand{\theoremname}{Theorem}

\providecommand{\claimname}{Claim}
\providecommand{\theoremname}{Theorem}

\providecommand{\claimname}{Claim}
\providecommand{\theoremname}{Theorem}

\providecommand{\claimname}{Claim}
\providecommand{\theoremname}{Theorem}

\providecommand{\claimname}{Claim}
\providecommand{\theoremname}{Theorem}

\providecommand{\claimname}{Claim}
\providecommand{\theoremname}{Theorem}

\makeatother

\providecommand{\claimname}{Claim}
\providecommand{\theoremname}{Theorem}

\begin{document}

\title{{\Huge{}{}{}{}{}{}On Stable Throughput of Cognitive Radio Networks
With Cooperating Secondary Users}}

\author{Kedar Kulkarni,~\IEEEmembership{Student~Member,~IEEE,} and Adrish
Banerjee,~\IEEEmembership{Senior Member,~IEEE}\thanks{The authors are with Department of Electrical Engineering, Indian
Institute of Technology Kanpur, Kanpur 208016, INDIA (e-mail: kulkarni@iitk.ac.in;
adrish@iitk.ac.in).}}
\maketitle
\begin{abstract}
In this paper, we study cooperative cognitive radio networks consisting
of a primary user and multiple secondary users. Secondary users transmit
only when primary user is sensed as silent and may interfere with
primary transmission due to imperfect sensing. When primary activity
is sensed correctly, secondary users cooperate with primary user by
assisting retransmission of failed packets of primary user. We analyze
packet throughput of primary and secondary users for three variations
of proposed cooperation method. Signal flow graph (SFG) based approach
is employed to obtain closed form expressions of packet throughput.
The analysis is done for two cases; individual sensing and cooperative
sensing. Further, we characterize optimal transmission probability
of secondary users that maximizes individual secondary packet throughput
keeping all queues in the system stable. Results present a comparison
of throughput performance of proposed cooperation methods under different
scenarios and show their benefits for both primary as well as secondary
user throughput.\end{abstract}


\section{Introduction}

Studies have shown that currently allocated wireless spectrum is highly
underutilized in temporal and spectral domain \cite{FCC}. Cognitive
radio (CR) is considered as a potential technology for efficient use
of allocated spectrum \cite{Haykin}. In cognitive radio systems,
unlicensed cognitive users, also called as secondary users (SUs) sense
the spectrum for activity of licensed users or primary users (PUs).
Depending on the sensing information, SUs make a decision to access
the spectrum for their own communication. Commonly used spectrum access
models are interweave mode and joint interweave-underlay mode. In
interweave mode, SUs access the spectrum only when PUs are sensed
to be silent. Due to errors in sensing, SU transmission may interfere
with PU transmission. In joint interweave-underlay mode, SUs transmit
even when PUs are sensed to be present. While opportunistically accessing
the spectrum allocated to PU, SUs must ensure that quality of service
(QoS) constraint of PUs is satisfied. Cognitive radio networks have
been extensively studied in literature under information-theoretic
framework with various objectives and constraints \cite{Goldsmith2,Devroye,Jovicic}.

In practice, data transmission is of bursty nature. Data arrives at
a transmitting source in form of packets of bits. In transmission,
a whole packet is lost if not decoded correctly at the receiver. Study
of such systems gives insights in network layer aspects like packet
throughput. Generally at a source, packets generated in upper layers
are stored in queues before transmission. In case of poor source-destination
link or high interference, packet transmission fails. If queue length
grows in unbounded manner, delivery of new packets cannot be guaranteed.
Such a system is said to be unstable. Thus, it is important to study
throughput of a stable system (called as stable throughput). In \cite{kedar3},
authors studied stable throughput tradeoff between only-interweave
and joint interweave-underlay mode for perfect sensing case. In \cite{Jeon,Duan},
authors analyzed effect of energy availability at each user on stable
throughput region of the system assuming perfect sensing. In \cite{Simeone_stable_throughput},
authors considered sensing errors and characterized optimal transmit
power of SU that keeps PU queue stable under interweave mode. In \cite{kedar3,Jeon,Duan,Simeone_stable_throughput},
users transmit without any cooperation between them. SUs may act as
relays for PU transmission and benefit due to cooperation as shown
in \cite{zhang_relay,Zou_relay,Lee_relay} under information theoretic
framework. In queue based systems, cooperation from SUs increases
packet throughput of PU. As a result PU packet queue is emptied more
often, providing more silent slots for SU to transmit. Shafie et al.
analyzed stable throughput of a three-node network with one node acting
as relay of finite queue size \cite{Shafie}. In \cite{krikidis},
authors analyzed stable throughput for a primary multi-access system
where SU receives packets from two PUs and relays them using superposition
coding technique when PU slot is idle. Fodor et al. studied tradeoff
between packet delay and energy consumption in a cooperative cognitive
network \cite{Fodor}. A common cooperation method is cooperative
relaying \cite{Kompella}. In cooperative relaying, SU receives unsuccessfully
transmitted PU packets and relays them to PU destination on next transmission
opportunity. An additional relay queue is needed at SU source for
this purpose. In \cite{Elmahdy}, authors considered a finite capacity
relay queue and proposed packet admission control at relay queue to
maximize SU packet throughput. Ashour et al. proposed admission control
as well as randomized service at relay queue and analyzed stable throughput
and packet delays \cite{Ashour}. Effect of energy availability on
cooperation and stable throughput was studied in \cite{kedar4} for
battery powered nodes.

\subsection{Main results}

In this paper, we propose a cooperation method where PU and SUs that
have received unsuccessful PU packets form a virtual multiple-input
single-output (MISO) system and retransmit the packet using distributed
orthogonal space-time block code (D-OSTBC). Depending on which SUs
assist PU transmission, we present three variations of the cooperation
method and analyze stable throughput of the system. We take into account
the effect of imperfect sensing. Only those SUs that sense presence
of PU correctly, can receive PU packet. Misdetecting SUs may interfere
with the packet transmission. Such cooperation model results in multiple
tradeoffs as large number of SUs provide higher cooperation to PU
but can potentially cause more interference to assisted transmission
due to imperfect sensing. Also, with increasing number of SUs, inter-SU
interference increases and negatively affects individual SU packet
throughput. Specifically, our contribution in this paper is as follows. 
\begin{itemize}
\item We propose and model the basic cooperation protocol where multiple
SUs that have received unsuccessful PU packet, retransmit the packet
along with PU using D-OSTBC. We then propose three versions of the
protocol, namely all relay cooperation (ARC), recurrent best relay
cooperation (R-BRC) and non-recurrent best relay cooperation (NR-BRC).
We analyze stable throughput of PU and SU for given protocols using
signal flow graph (SFG) theory. 
\item The stable throughput analysis is done for two cases; individual sensing
(IS) and cooperative sensing (CS). In individual sensing each SU senses
spectrum and takes decision to transmit independently. In cooperative
sensing, SUs share their sensing decisions (hard or soft decisions)
and take a collective decision on availability of spectrum. 
\item We then characterize optimal transmission probability of SU that maximizes
individual SU packet throughput while keeping PU queue stable. 
\item Finally, we present numerical results to study various tradeoffs arising
with different number of SUs, varying SU transmit power and sensing
type (individual or cooperative). 
\end{itemize}

\subsection{Related work\label{sub:Related-work}}

In \cite{Kompella,Elmahdy,Ashour,kedar4}, stable throughput with
cooperative relaying was studied under joint interweave-underlay mode
for a simple two-user model. Fanous et al. extended the model for
the case of multiple SUs where SUs receive unsuccessful PU packet
and relay it using D-OSTBC in the next silent slot \cite{fanous}.
Due to joint interweave-underlay mode in \cite{Kompella}, there is
a chance that a PU packet being relayed by SU may collide with new
packet transmitted by PU. Model in \cite{fanous} solved this issue
by mandating that SUs relay PU packets only in silent slots. However
it assumed perfect sensing. In contrast, we consider effect of imperfect
sensing in this paper. Further, in cooperative relaying, a packet
is removed from PU queue if it is successfully received at PU destination
or at any of the SU sources. If a PU packet is received by a SU, responsibility
of delivering the packet lies solely at SU. Such cooperation is ineffective
if link between SU sources and PU destination is weak. Cooperation
method proposed in this paper resolves this issue as a packet is removed
from PU queue only when it is received successfully at PU destination.
Thus, packet departure rate of PU queue equals goodput of PU, unlike
cooperation method in \cite{fanous}.

Canzian et al. \cite{Badia} studied throughput and average time for
packet delivery in a non-cognitive network for two cooperation scenarios--
forced cooperation where best relay retransmits a failed packet and
voluntary cooperation where a user may act as relay to get higher access
probability in return. In our paper, due to presence of PU, relaying
capability of SUs is affected by sensing errors and interference by
other users. In \cite{Badia}, authors modelled packet transmission
process of automatic repeat request (ARQ) mechanism as Markov chain
assuming that at most one retransmission per packet is allowed. We
model transmission process of proposed protocols in a similar way
but the signal flow graph approach employed for throughput analysis puts
no restriction on number of retransmissions.

\subsection{Organization}

In Section II, we present the system model. In Section III, proposed
cooperation method and its variations are explained. Also packet throughput
expressions for given methods are derived. In Section IV and V, packet
throughput of PU and SU is analyzed for individual sensing case and
cooperative sensing case respectively. In Section VI, we present numerical
results to demonstrate performance of proposed methods. Finally we
conclude in Section VII.

\section{System model}

\begin{figure}
\centering

\includegraphics[scale=0.25]{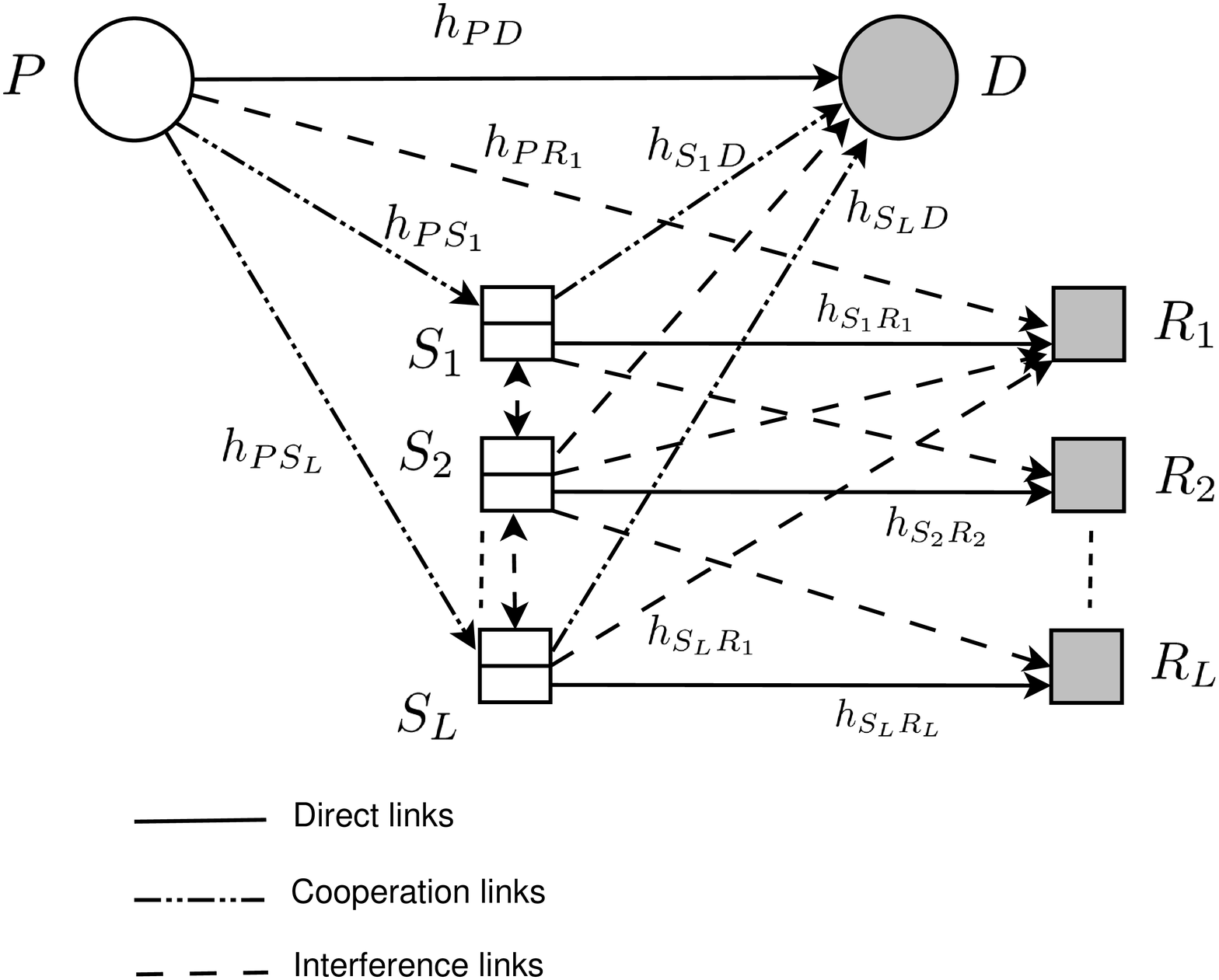}

\protect\protect\protect\protect\protect\caption{System model for CR network with cooperating secondary users\label{fig:System-model}}
\vspace{-0.2cm}
\end{figure}

As shown in Fig. \ref{fig:System-model}, the system consists of a
PU transmitting from source $P$ to destination $D$. A secondary
network of $L$ SUs uses the same frequency band as that of PU to
transmit packets from source $S_{j}$ to respective destination $R_{j}$,
$j\in\left\{ 1,\,2,\dots,\,L\right\} $. All users store packets in
queues before transmission. Each SU source has one queue for its own
packets and one relay queue to store unsuccessful packets of PU. All
queues have infinite storing capacity \cite{fanous,Fodor}. This is
a good approximation for practical systems with large queue sizes
as packet loss probability due to buffer overflow is low. Similar
to \cite{fanous}, we assume that all SUs have backlogged packet queues,
that is each SU always has packets to transmit. Time is slotted such
that duration of a slot equals time required to transmit one packet.
Packet arrival at PU queue is a stationary Bernoulli process with
average rate $\lambda_{P}\in\left[0,\,1\right]$ packets per slot.
Average packet departure rates of PU and SUs are denoted by $\mu_{P}$
and $\mu_{S_{j}},\,j\in\left\{ 1,\,2,\dots L\right\} $ respectively.

\subsection{Stable throughput}

A queue is said to be stable if the queue length does not increase
in unbounded manner. Unlike an unstable queue where packets may get
queued up indefinitely, packet delivery is guaranteed in a stable
queue. For a system to be stable, all queues in the system should
be stable. Packet throughput achieved in such a system is called stable
throughput. Thus, for the system of Fig. \ref{fig:System-model} to
be stable, we need PU queue as well as relay queues at all SUs to
be stable. In proposed cooperation protocol, relay queues at SUs are
always stable due to the fact that relay queue length cannot exceed
$1$ as explained later in Section III. The PU queue evolves with
time as

\[
Q_{P}^{t+1}=\left[Q_{P}^{t}-Y_{P}^{t}\right]^{+}+X_{P}^{t},
\]
where $Q_{P}^{t}$ denotes length of PU queue at the beginning of
time slot $t$, $Y_{P}^{t}$ is number of packet departures in time
slot $t$ and $X_{P}^{t}$ denotes number of arrivals in time slot
$t$. The operation $\left[x\right]^{+}$ denotes $\max\left(0,\,x\right)$.
PU queue is stable if for every $x\in\mathbb{N}_{0}$, $\lim_{t\rightarrow\infty}\Pr\left[Q_{P}^{t}<x\right]=F\left(x\right)$
with $\lim_{x\rightarrow\infty}F\left(x\right)=1$. We use Loynes'
criteria for queue stability which states that, for jointly stationary
packet arrival and departure processes, the queue is stable if average
packet arrival rate is less than average packet departure rate \cite{loynes_stability}.
Packet departure depends on channel fading, interference by SUs as
well as cooperation offered by SUs, and is independent of packet arrivals.
As PU packet departure rate is given by $\mu_{P}=E\left[Y_{P}^{t}\right]$,
condition for PU queue stability is 
\[
\lambda_{P}<\mu_{P}.
\]
Thus, to keep system stable, SUs should control the interference caused
to PU and ensure PU queue stability.

\subsection{Physical layer model}

All channels are independent Rayleigh block fading, that is, channel
coefficients remain constant in one slot and change independently
from slot to slot. Channel coefficient of link between source $s\in\left\{ P,\,S_{1},\,S_{2},\dots,\,S_{L}\right\} $
and destination $d\in\left\{ D,\,R_{1},\,R_{2},\dots,\,R_{L},\,S_{1},\,S_{2},\dots,\,S_{L}\right\} $
is $h_{sd}\sim\mathcal{CN}\left(0,\,\sigma_{sd}^{2}\right)$. We can
classify the communication links in $6$ classes; PU source to PU
destination, PU source to SU destinations, PU source to SU sources,
SU sources to SU destinations, SU sources to PU destination and SU
source to other SU sources. Fig. \ref{fig:System-model} shows direct
links between a source and its intended destination by continuous
lines, interference links by dashed lines and cooperation links by
dash-and-dotted lines. As done in \cite{fanous}, for mathematical
tractability of our analysis, we consider a symmetric case where distribution
parameters of all channels in one class are same. This is possible
if all SU source nodes are in close vicinity and all SU destinations
are in close vicinity. The proposed framework for analysis of stable
throughput remains unchanged for a general asymmetric case. For the
symmetrical case, channel properties are as follows. 
\begin{itemize}
\item PU source to PU destination $h_{PD}\sim\mathcal{CN}\left(0,\,\sigma_{PD}^{2}\right)$, 
\item PU source to $i$th SU destination $h_{PR_{i}}\sim\mathcal{CN}\left(0,\,\sigma_{PR}^{2}\right)$
for $i\in\left\{ 1,\,2,\dots,\,L\right\} $, 
\item PU source to $i$th SU source $h_{PS_{i}}\sim\mathcal{CN}\left(0,\,\sigma_{PS}^{2}\right)$
for $i\in\left\{ 1,\,2,\dots,\,L\right\} $, 
\item $i$th SU source to $j$th SU destination $h_{S_{i}R_{j}}\sim\mathcal{CN}\left(0,\,\sigma_{SR}^{2}\right)$
for $i,\,j\in\left\{ 1,\,2,\dots,\,L\right\} $, 
\item $i$th SU source to PU destination $h_{S_{i}D}\sim\mathcal{CN}\left(0,\,\sigma_{SD}^{2}\right)$
for $i\in\left\{ 1,\,2,\dots,\,L\right\} $, 
\item $i$th SU source to $j$th SU source $h_{S_{i}S_{j}}\sim\mathcal{CN}\left(0,\,\sigma_{SS}^{2}\right)$
for $i\in\left\{ 1,\,2,\dots,\,L\right\} $, $j\in\left\{ 1,\,2,\dots,\,L\right\} /\{i\}$. 
\item Noise is additive white Gaussian (AWGN) $n\sim\mathcal{CN}\left(0,\,\sigma_{N}^{2}\right)$. 
\end{itemize}

\subsection{Spectrum sensing and access}

PU and SUs follow time slotted synchronous communication \cite{Zhao}.
The slot is divided in two parts; first of duration $T_{s}$ allocated
to tasks like spectrum sensing and exchange of control information,
second of duration $T_{t}$ used for transmission of packets. In the
first part, PU sends pilot signals to PU destination while SUs employ
pilot based spectrum sensing to detect presence of PU transmission
\cite{PilotSense_Wang}. Due to channel fading, two types of sensing
errors may occur, namely misdetection and false alarm. Misdetection
happens when PU is active but a SU senses it as idle. False alarm
occurs when PU is sensed as active when in fact it is silent. We denote
probabilities of correct detection and false alarm by $p_{d}$ and
$p_{f}$ respectively. Due to independence of fading channels between
PU source and SU source nodes, misdetection and false alarm events
are independent for all SUs. Further, due to the assumption of symmetry,
values of $p_{d}$ and $p_{f}$ are same for all SUs. In Section V,
we also consider the case of cooperative sensing where SUs share their
sensing data (hard sensing data or soft sensing data) and a collective
decision is taken on the availability of spectrum \cite{Peh,Ghasemi,Akyildiz}.
Probabilities of correct detection and false alarm in this case are
denoted as $p_{d}^{*}$ and $p_{f}^{*}$ respectively. In case of
cooperative sensing, if PU is sensed as active, all SUs keep silent.

In the transmission duration, PU transmits using a feedback mechanism
where destination node $D$ sends an acknowledgement (ACK) to source
$P$ when a packet is correctly received. If packet transmission is
unsuccessful, a negative acknowledgement (NACK) for that packet is
sent. We make following assumptions regarding the system model as
done in \cite{Kompella}. 
\begin{itemize}
\item The feedback channel is an error-free broadcast channel. Thus, SUs
can overhear ACKs and NACKs sent by PU destination $D$. 
\item Feedback is available immediately after packet transmission. 
\item SUs are able to receive packets transmitted by PU. This is possible
if the SU sources lie in vicinity of PU source transmitting with an
omni-directional antenna. 
\item SU can either receive or transmit at a time but cannot do both actions
simultaneously. This holds true in most practical cases where nodes
are equipped with a single transreceiver pair. Thus, a SU is able to
receive PU packets only when it is silent. 
\end{itemize}
If PU is sensed as silent, in transmission duration, each SU transmits
independently with probability $q$. If PU is sensed as present, a
SU keeps silent (interweave mode) and cooperates using methods described
in Section III. Transmission of SU's own packets may occur in two
cases; one when PU is active and SU misdetects, other when PU is silent
and there is no false alarm. Usually the target detection probability
is high, hence second case dominates achievable throughput \cite{Liang_zheng}.
Also probability of successful packet transmission of SU is less in
case of interference from PU due to generally high PU transmit power.
Thus, we restrict analysis of SU packet throughput to the second case
which gives a lower bound on SU packet throughput performance. The
bound is tight when PU transmit power is very high and link between
PU source to SU destination is strong. The lower bound and actual
performance coincide in case of perfect sensing, as PU and SUs do
not interfere.

\subsection{Probability of successful packet transmission\label{sub:Probability-of-successful}}

PU transmits with power $P_{P}$ and each SU transmits with power
$P_{S}$. PU and SU packets have fixed length of $\mathcal{B}$ bits.
A packet is delivered successfully to the intended receiver in a slot
if instantaneous channel capacity on the source-destination link is
greater than $\frac{\mathcal{B}}{T_{t}}$ bits/s. Instantaneous channel
capacity can be given as $\mathcal{R}=W\log_{2}\left(1+\gamma\right)$
bits/s where $\gamma$ is received signal to interference plus noise
ratio (SINR) and $W$ is channel bandwidth. Then the probability of
successful packet transmission is $\Pr\left[\gamma>2^{\frac{\mathcal{B}}{WT_{t}}}-1\right].$
In rest of the paper, we use the notation $\beta=2^{\frac{\mathcal{B}}{WT_{t}}}-1$.

\section{Cooperation methods and signal flow graph representation}

If a SU correctly detects PU, it can receive packet transmitted by
PU. This enables SUs to cooperate with PU as follows. If the PU packet
transmission is unsuccessful, a NACK is sent by PU destination which
can be heard by all SUs. Upon receiving NACK, a SU puts the PU packet
in its relay queue, provided the packet was correctly received at
the SU source. In retransmission phase, multiple sources including
the PU source, have the same packet ready to be transmitted. Each
source can act as an antenna in a MISO channel and transmit the packet
using orthogonal space-time block coding (OSTBC) scheme \cite{Tarokh_OSTBC,fanous}.
In OSTBC, the packet is encoded in blocks of bits which are distributed
among different antennas and across time. For this purpose, all transmitting
sources need channel state information (CSI) of other transmitting
sources. Also each transmitting source must know which antenna it
mimics in the virtual MISO, in order to transmit appropriate parts
of the packet according to the corresponding space-time matrix. This
can be achieved by coordination between the sources on a low bandwidth
control channel or by prior indexing. In the next slot, PU as well
as all the cooperating SUs retransmit the PU packet using distributed-OSTBC.
Diversity results in higher SINR at PU receiver and probability of
successful packet transmission increases. If an ACK is received after
the retransmission, the packet is removed from PU queue and from relay
queues of cooperating SUs. If a NACK is received, SUs and PU keep
retransmitting the same packet until it is received successfully at
PU destination. SUs which don't have the PU packet continue to sense
the spectrum and may interfere with assisted retransmission in case
of misdetection. It should be noted that priority is given to PU packets,
that is SUs always transmit from relay queue if the relay queue is
non-empty. As PU does not transmit a new packet until the retransmitted
packet is correctly received, there is at most 1 packet in any relay
queue.

A case may arise where assisted retransmission is unsuccessful and
some of the \textit{listening} SUs receive the PU packet from assisted
retransmission. Thus, more SUs can cooperate with PU in the next slot.
However, due to multiple queue interactions between relay queues,
keeping track of how many SUs receive packets from assisted retransmissions
is complicated. Thus, we restrict our analysis to the case where same
group of assisting SUs participates in retransmission if current retransmission
is unsuccessful-- that is, other SUs don't receive packets from assisted
retransmission. This gives us a lower bound on PU packet throughput
performance.

\subsection{Methods of cooperation}

We analyze stable throughput for three variations of the proposed
cooperation method. 
\begin{enumerate}
\item All Relay Cooperation (ARC) - All SUs that receive unsuccessful PU
packet transmit in retransmission phase. 
\item Recurrent Best Relay Cooperation (R-BRC) - Out of all assisting SUs,
only the SU that has highest instantaneous channel gain on SU source
to PU destination link participates in retransmission. We refer to
such a SU as ``best'' SU relay. Other assisting SUs remain silent
and retain the packet. If retransmission is unsuccessful, best SU
relay selection is repeated and new best SU assists PU transmission.
Note that the same SU may be chosen as best SU relay again depending
on the instantaneous channel gains. Best SU relay selection can be
done in duration $T_{s}$ using time-out timers inversely proportional
to channel gains as done in \cite{Ezzeldin_BestRelay}. 
\item Non-recurrent Best Relay Cooperation (NR-BRC) - In this method, unlike
R-BRC, other SUs discard PU packet after best SU relay selection is
performed. If retransmission is unsuccessful, the same SU assists
irrespective of whether it has the best SU source to PU destination
channel. Other SUs continue sensing and accessing the spectrum with
probability $q$.
\end{enumerate}

\subsection{Signal flow graph representation}

In commonly used automatic repeat request (ARQ) protocols, source
node attempts retransmission of an unsuccessful packet until ACK for
the packet is received. Packet transmission process of ARQ protocols
can be represented by signal flow graphs (SFG). SFG representation
and subsequent graph reduction provides an efficient way to analyze
packet throughput \cite{Lu,Lu2,Nosratinia}. To calculate PU packet
throughput, we first represent the cooperation methods by signal flow
graphs.

\begin{figure}
\centering

\includegraphics[scale=0.15]{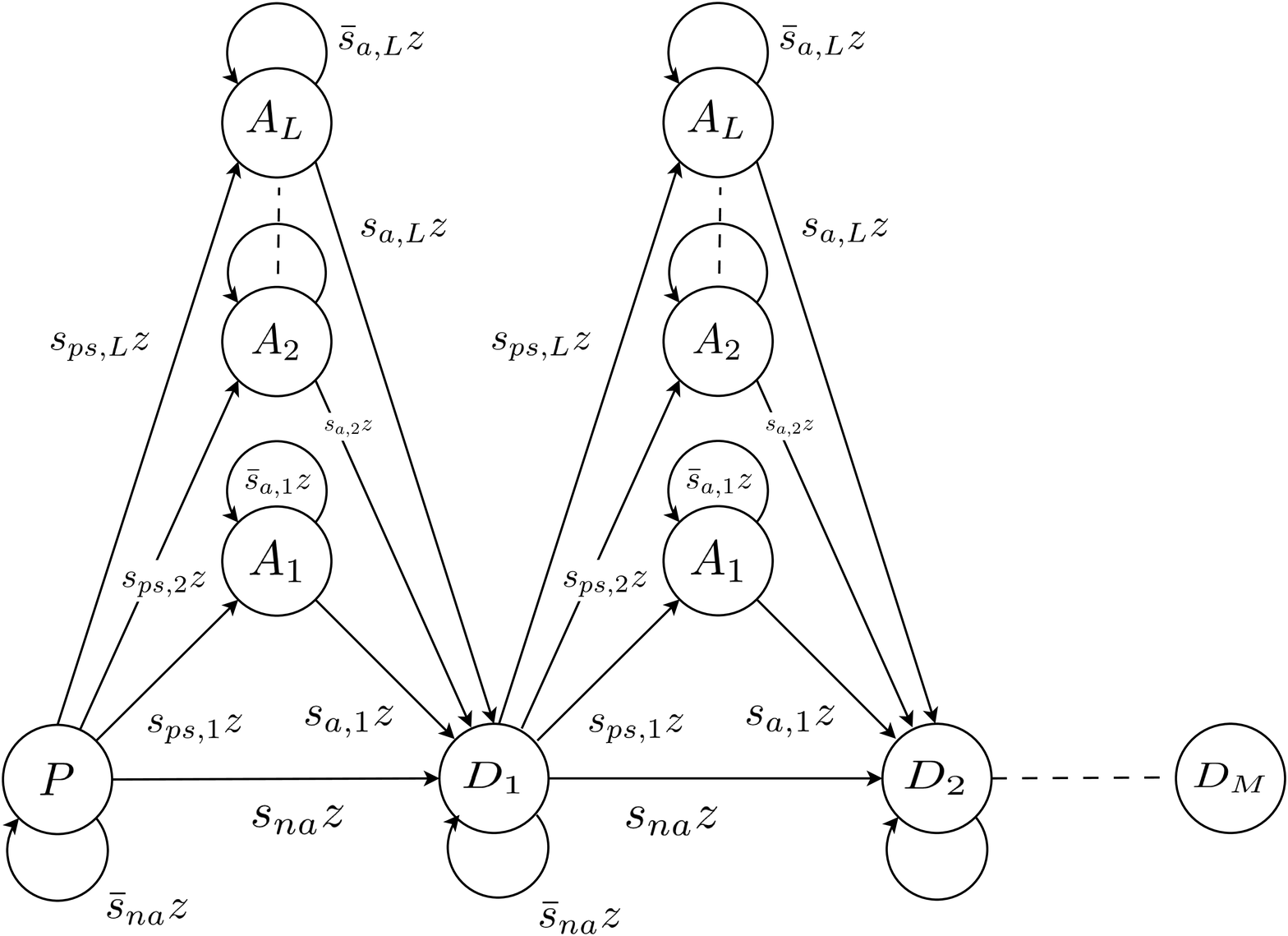}

\protect\protect\protect\caption{Signal flow graph for PU transmission under ARC/R-BRC method\label{fig:Signal-flow-graph}}
\end{figure}

\subsubsection{ARC/R-BRC}

Fig. \ref{fig:Signal-flow-graph} shows SFG of ARC and R-BRC methods
for transmission of $M$ PU packets. We use $z$ as one slot length
operator. PU packets originate from input $P$. State $D_{j},\,j\in\left\{ 1,\,2,\dots,\,M\right\} $
represents the state where ACK for $j$th packet is received and transmission
of new packet begins. Transmission of $M$ packets is over at output
$D_{M}$. Any path leading from $P$ to $D_{M}$ corresponds to successful
transmission of $M$ packets. State $A_{n}$ represents the assist state where
assistance from $n$ SUs is possible for retransmission of a packet.

We first discuss transmission of a single packet from input $P$ to
output $D_{1}$. A packet is successfully transmitted without assistance
from SU with probability $s_{na}$, indicated by link $P-D_{1}$.
With probability $s_{ps,n}$, PU transmission is unsuccessful and
the packet is received by $n$ SUs as indicated by the link $P-A_{n}$.
As there are $L$ SUs in the system, there can be at most $L$ assist
states. Self loop at $P$ shows that PU transmission is unsuccessful
and no SU is able to receive the unsuccessful packet. This happens
with probability $\bar{s}_{na}$. Thus, we have 
\begin{equation}
s_{na}+\bar{s}_{na}+\sum_{n=1}^{L}s_{ps,n}=1.\label{eq:SFG1}
\end{equation}

When $n$ SUs assist, a packet retransmission is successful with probability
$s_{a,n}$ as shown by $A_{n}-D_{1}$ link. Probability $s_{a,n}$
depends on whether all assisting SUs transmit (in ARC) or only the
best SU transmits (in R-BRC). Self loop at $A_{n}$ indicates that
assisted transmission is unsuccessful. This happens with probability
$\bar{s}_{a,n}$. As we assume that non-assisting SUs do not receive
packets from assisted retransmissions, there is no link between two
assist states $A_{n}-A_{m},\,n\neq m$. Then we have 
\begin{equation}
s_{a,n}+\bar{s}_{a,n}=1\,\,\mbox{for}\,n\in\left\{ 1,\,2,\dots,\,L\right\} .\label{eq:SFG2}
\end{equation}
The links between $P-D_{1}$ are repeated for transmission of $M$
packets. 
\begin{claim}
PU packet throughput for cooperation protocol in Fig. \ref{fig:Signal-flow-graph}
is given by 
\begin{equation}
\mu_{P}=\left(1-\bar{s}_{na}\right)\left[1+\sum_{n=1}^{L}\frac{s_{ps,n}}{s_{a,n}}\right]^{-1}.\label{eq:thru1}
\end{equation}
\end{claim}
\begin{IEEEproof}
We first determine transfer function of the graph in Fig. \ref{fig:Signal-flow-graph}.
For single packet transmission from $P$ to $D_{1}$, there are $\left(L+1\right)$
parallel forward paths as listed below. 
\begin{align*}
 & \mbox{Forward path }F_{1}\,:\,\,P-A_{1}-D_{1}\\
 & \mbox{Forward path }F_{2}\,:\,\,P-A_{2}-D_{1}\\
 & \qquad\vdots\\
 & \mbox{Forward path }F_{L}\,:\,\,P-A_{L}-D_{1}\\
 & \mbox{Forward path }F_{L+1}:\,\,P-D_{1}
\end{align*}
Each path $F_{n},\,n=1,\,2,\dots,\,L$ has path gain $G_{F_{n}}=s_{ps,n}s_{a,n}z^{2}$
and it touches a loop at $A_{n}$ with loop gain $L_{F_{n}}=\bar{s}_{a,n}z$.
Path $F_{L+1}$ has path gain $G_{F_{L+1}}=s_{na}z$. By Mason's gain
formula \cite{Mason}, we can write transfer function of each forward
path as
\begin{align}
H_{F_{n}}\left(z\right) & =\frac{s_{ps,n}s_{a,n}z^{2}}{1-\bar{s}_{a,n}z}\,\,\mbox{for}\,\,n=1,\,2,\dots,\,L,\label{eq:forward1}\\
H_{F_{L+1}}\left(z\right) & =s_{na}z\,.\label{eq:forward2}
\end{align}

We can replace $\left(L+1\right)$ parallel branches connecting two
nodes in the same direction by a single branch with path gain equal
to sum of path gains of the parallel branches. After merging the parallel
branches using (\ref{eq:forward1}) and (\ref{eq:forward2}), we get
a single forward path from $P$ to $D_{1}$ with gain 
\[
G_{F}=s_{na}z+\sum_{n=1}^{L}\frac{s_{ps,n}s_{a,n}z^{2}}{1-\bar{s}_{a,n}z}.
\]
The path touches a self-loop at $P$ having loop gain $L_{F}=\bar{s}_{a,n}z$.
Then using Mason's gain formula, transfer function for transmission
of single packet from $P$ to $D_{1}$ is given by 
\[
H\left(z\right)=\frac{s_{na}z+\sum_{n=1}^{L}\frac{s_{ps,n}s_{a,n}z^{2}}{1-\bar{s}_{a,n}z}}{1-\bar{s}_{na}z}.
\]
For transmission of $M$ packets, there are $M$ such branches in
series. Thus, overall transfer function from input $P$ to output $D_{M}$
is 
\begin{equation}
H_{M}\left(z\right)=\left[\frac{s_{na}z+\sum_{n=1}^{L}\frac{s_{ps,n}s_{a,n}z^{2}}{1-\bar{s}_{a,n}z}}{1-\bar{s}_{na}z}\right]^{M}.\label{eq:transfer_fn}
\end{equation}
Average number of slots required to transmit $M$ packets is given
by $H_{M}^{'}\left(1\right)=\left.\frac{dH_{M}\left(z\right)}{dz}\right|_{z=1}$
\cite{Lu}. Thus, we can write PU packet throughput in packets/slot
as 
\begin{equation}
\mu_{P}=\lim_{M\rightarrow\infty}\frac{M}{\left.\frac{dH_{M}\left(z\right)}{dz}\right|_{z=1}}.\label{eq:PU_thru1}
\end{equation}
After solving (\ref{eq:PU_thru1}) using (\ref{eq:SFG1}) and (\ref{eq:SFG2}),
we get 
\begin{equation}
\mu_{P}=\frac{1-\bar{s}_{na}}{1+\sum_{n=1}^{L}\frac{s_{ps,n}}{s_{a,n}}}.\label{eq:PU_thru2}
\end{equation}

\end{IEEEproof}
\begin{figure}
\centering

\includegraphics[scale=0.15]{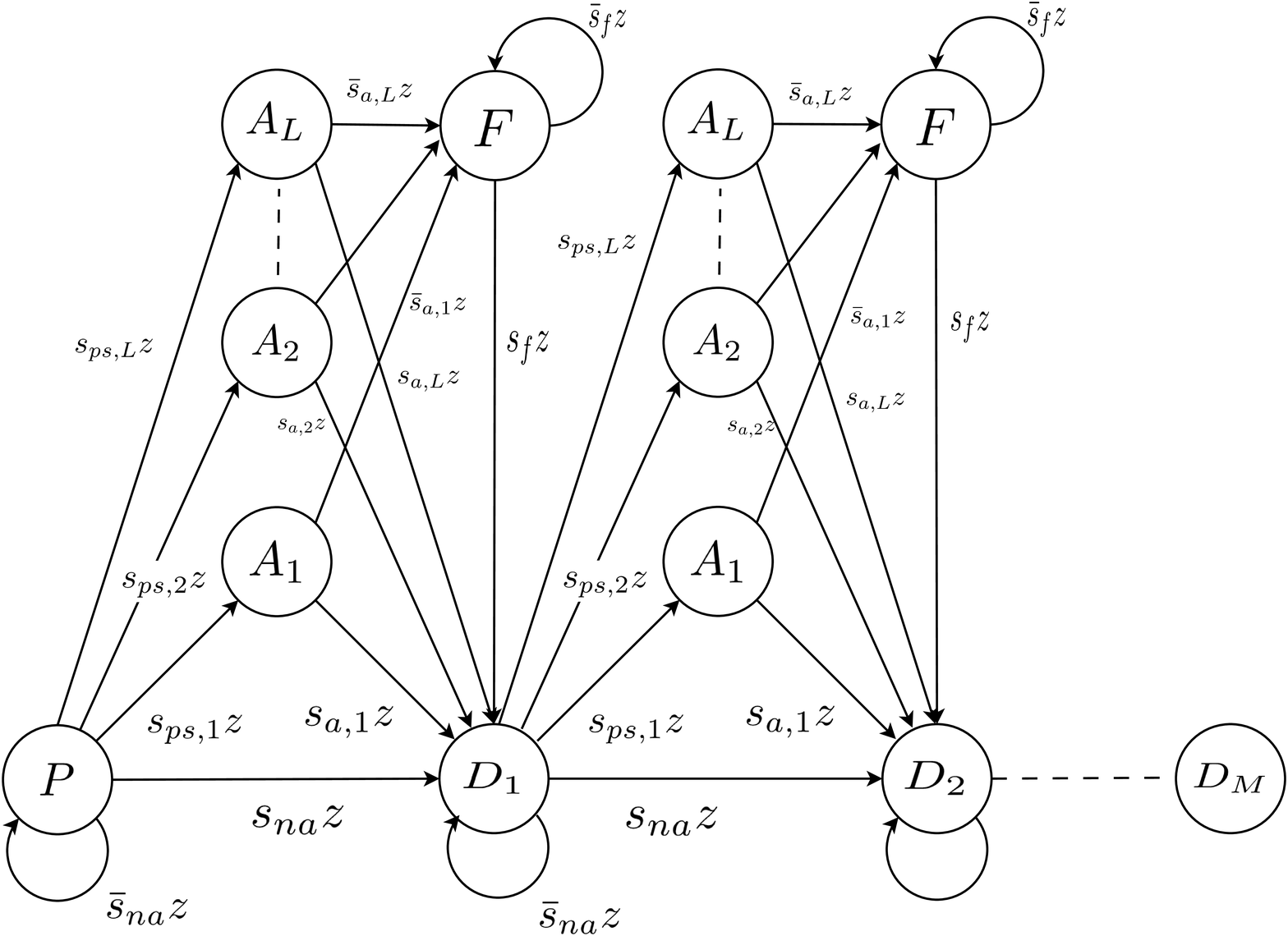}

\protect\protect\protect\caption{Signal flow graph for PU transmission under NR-BRC method\label{fig:Signal-flow-graph-BRC}}
\end{figure}

\subsubsection{NR-BRC}

Fig. \ref{fig:Signal-flow-graph-BRC} shows SFG of NR-BRC method.
As explained in the case of ARC/R-BRC, (\ref{eq:SFG1}) and (\ref{eq:SFG2})
hold true for NR-BRC. But unlike R-BRC, best relay selection is not
performed again if assisted transmission is unsuccessful. Thus, if
retransmission is unsuccessful, process does not return to the same
assist state. Instead it goes to a ``fresh attempt'' state denoted
by $F$ where the same SU assists PU retransmission irrespective of
whether it is the best SU or not. Packet transmission from fresh attempt
state is successful with probability $s_{f}$ as shown by $F-D_{1}$
link. With probability $\bar{s}_{f}$, the retransmission is unsuccessful
and process remains in the same state. Thus, we have 
\begin{equation}
s_{f}+\bar{s}_{f}=1.\label{eq:SFG3}
\end{equation}

\begin{claim}
PU packet throughput for cooperation protocol in Fig. \ref{fig:Signal-flow-graph-BRC}
is given by 
\begin{equation}
\mu_{P}=\left(1-\bar{s}_{na}\right)\left[1+\sum_{n=1}^{L}s_{ps,n}+\sum_{n=1}^{L}\frac{s_{ps,n}\bar{s}_{a,n}}{s_{f}}\right]^{-1}.\label{eq:thru2}
\end{equation}
\end{claim}
\begin{IEEEproof}
We first list all forward paths for single packet transmission from
$P$ to $D_{1}$. 
\begin{align*}
 & \mbox{Forward path }F_{1}\,:\,\,P-A_{1}-D_{1}\\
 & \mbox{Forward path }F_{2}\,:\,\,P-A_{2}-D_{1}\\
 & \qquad\vdots\\
 & \mbox{Forward path }F_{L}\,:\,\,P-A_{L}-D_{1}\\
 & \mbox{Forward path }F_{L+1}\,:\,\,P-A_{1}-F-D_{1}\\
 & \mbox{Forward path }F_{L+2}\,:\,\,P-A_{2}-F-D_{1}\\
 & \qquad\vdots\\
 & \mbox{Forward path }F_{2L}\,:\,\,P-A_{L}-F-D_{1}\\
 & \mbox{Forward path }F_{2L+1}\,:\,\,P-D_{1}
\end{align*}
There are two self-loops; one at input $P$ with loop gain $L_{1}=\bar{s}_{na}z$
and one at state $F$ with loop gain $L_{2}=\bar{s}_{f}z$. As both
loops are non-touching, graph determinant $\Delta$ is given as 
\begin{equation}
\Delta=1-L_{1}-L_{2}+L_{1}L_{2}=1-\bar{s}_{na}z-\bar{s}_{f}z+\bar{s}_{na}\bar{s}_{f}z^{2}.\label{eq:determinant}
\end{equation}
Using (\ref{eq:determinant}), we get path gains and co-factors associated
with each forward path as 
\[
G_{F_{n}}=\begin{cases}
s_{ps,n}s_{a,n}z^{2} & \!\!\mbox{for}\,\,n=1,\,2,\dots,\,L\\
s_{ps,n-L}\bar{s}_{a,n-L}s_{f}z^{3} & \!\!\mbox{for}\,\,n=L+1,\,L+2,\dots,\,2L\\
s_{na}z & \!\!\mbox{for}\,\,n=2L+1
\end{cases},
\]
\[
\Delta_{F_{n}}=\begin{cases}
1-\bar{s}_{f}z & \,\,\mbox{for}\,\,n=1,\,2,\dots,\,L\\
1 & \,\,\mbox{for}\,\,n=L+1,\,L+2,\dots,\,2L\\
1-\bar{s}_{f}z & \,\,\mbox{for}\,\,n=2L+1
\end{cases}.
\]
Using Mason's gain formula, transfer function of $P-D_{1}$ link is
given by 
\[
H\left(z\right)=\frac{\sum_{n=1}^{2L+1}G_{F_{n}}\Delta_{F_{n}}}{\Delta}
\]
For transmission of $M$ packets, we have $H_{M}\left(z\right)=\left[H\left(z\right)\right]^{M}$.
Using same approach followed in ARC/R-BRC case, we get 
\begin{equation}
\mu_{P}=\frac{1-\bar{s}_{na}}{1+\sum_{n=1}^{L}s_{ps,n}+\sum_{n=1}^{L}\frac{s_{ps,n}\bar{s}_{a,n}}{s_{f}}}.
\end{equation}

\end{IEEEproof}
In the next section, we derive values of probabilities used in (\ref{eq:thru1})
and (\ref{eq:thru2}). This allows us to calculate PU and SU packet
throughput.

\begin{table*}
\centering \setlength{\extrarowheight}{8pt}

\protect\caption{Notation for probabilities of successful packet transmission\label{tab:Notations-for-probabilities}}

\begin{tabular}{|c|c|}
\hline 
Notation  & Probability of successful transmission\tabularnewline
\hline 
$\mathcal{U}_{a}\left(n,m\right)$  & PU source to PU destination when \textbf{$n$ SUs assist and $m$
SUs interfere}\tabularnewline
$\mathcal{U}_{b}\left(n,m\right)$  & PU source to PU destination when \textbf{best SU out of $n$ SUs assists
and $m$ SUs interfere}\tabularnewline
$\mathcal{V}\left(m\right)$  & SU source to SU destination when PU doesn't interfere and $m$ SUs
interfere\tabularnewline
$\mathcal{W}\left(m\right)$  & PU source to SU source when $m$ SUs interfere\tabularnewline
\hline 
\end{tabular}
\end{table*}

\section{Throughput analysis - Individual sensing (IS)}

We define the notation used to denote probabilities of successful packet
transmission under various cases in Table \ref{tab:Notations-for-probabilities}.
Exact closed form expressions for $\mathcal{U}_{a}\left(n,m\right)$
and $\mathcal{U}_{b}\left(n,m\right)$ are derived in Appendix A and
Appendix B. Expressions of $\mathcal{W}\left(m\right)$ and $\mathcal{V}\left(m\right)$
can be derived as special cases of $\mathcal{U}_{a}\left(n,m\right)$
and have been given in \cite[Eq.(11), (17)]{fanous}. Using results
in (\ref{eq:thru1}) and (\ref{eq:thru2}), we now analyze packet
throughput of PU and SU for the case where each SU senses individually
and transmits based on own sensing decision.

\subsection{Primary user throughput}

\subsubsection{ARC\label{sub:ARC_analyze}}

When PU is present, a non-assisting SU interferes with PU in case
of misdetection, that is with probability $\left(1-p_{d}\right)q$.
When $m$ SUs transmit and no SU assists, PU transmission is successful
with probability $\mathcal{\mathcal{U}}_{a}\left(0,m\right)$. Then
probability of successful transmission of PU packet without any assistance
is 
\[
s_{na}=\sum_{m=0}^{L}\binom{L}{m}\,\left[\left(1-p_{d}\right)q\right]^{m}\left[1-\left(1-p_{d}\right)q\right]^{L-m}\mathcal{U}_{a}\left(0,m\right).
\]
Substituting value of $\mathcal{U}_{a}\left(0,m\right)$ from (\ref{eq:Qd_no_assist}),
we get 
\begin{equation}
s_{na}=\exp\left(\frac{-\sigma_{N}^{2}\beta}{P_{P}\sigma_{PD}^{2}}\right)\left[1-\frac{B_{P}\left(1-p_{d}\right)q}{1+B_{P}}\right]^{L},\label{eq:Sna}
\end{equation}
where $B_{P}=\frac{\beta P_{S}\sigma_{SD}^{2}}{P_{P}\sigma_{PD}^{2}}$.

SUs that correctly detect PU remain silent and try to receive PU packet.
Misdetecting SUs in the vicinity may cause interference at the receiving
SU sources. Let $\mathcal{O}_{PD}$ be the event that PU transmission
on $P-D$ link is unsuccessful. Also let $\overline{\mathcal{O}_{PS}\left(n\right)}$
be the event that $n$ SUs receive PU packet. Then overall probability
that direct transmission of PU is unsuccessful but $n$ SUs are able
to receive PU packet is given by 
\begin{align*}
s_{ps,n} & =\sum_{l=n}^{L}\Pr\left\{ l\,\mbox{out of}\,L\,\mbox{SUs detect PU}\right\} \\
 & \times\sum_{m=0}^{L-l}\Pr\left\{ m\,\mbox{SUs out of}\,\left(L-l\right)\,\mbox{interfere}\right\} \\
 & \times\Pr\left\{ \mathcal{O}_{PD}\cap\overline{\mathcal{O}_{PS}\left(n\right)}\,|\,m\,\mbox{SUs interfere}\right\} .
\end{align*}

Given that $l$ SUs try to receive PU packet, probability that $n\leq l$
SUs actually receive the packet under interference from $m$ SUs is
$\binom{l}{n}\left[\mathcal{W}\left(m\right)\right]^{n}\left[1-\mathcal{W}\left(m\right)\right]^{l-n}$.
Noting that events $\mathcal{O}_{PD}$ and $\overline{\mathcal{O}_{PS}\left(n\right)}$
are independent, we can write

\begin{align}
s_{ps,n} & =\sum_{l=n}^{L}\binom{L}{l}\,p_{d}^{l}\left(1-p_{d}\right)^{L-l}\sum_{m=0}^{L-l}\binom{L-l}{m}\nonumber \\
 & \qquad\times q^{m}\left(1-q\right)^{L-l-m}\left[1-\mathcal{U}_{a}\left(0,m\right)\right]\nonumber \\
 & \qquad\times\binom{l}{n}\left[\mathcal{W}\left(m\right)\right]^{n}\left[1-\mathcal{W}\left(m\right)\right]^{l-n}.\label{eq:Spsn}
\end{align}

All SUs that receive PU packet assist with the packet transmission.
Remaining SUs that don't have the PU packet, continue sensing spectrum
and may interfere in transmission duration in case of misdetection.
Thus, probability of successful PU transmission when $n$ SUs assist
is 
\begin{align}
s_{a,n} & =\sum_{m=0}^{L-n}\binom{L-n}{m}\,\left[\left(1-p_{d}\right)q\right]^{m}\nonumber \\
 & \qquad\times\left[1-\left(1-p_{d}\right)q\right]^{L-n-m}\mathcal{U}_{a}\left(n,m\right).\label{eq:San}
\end{align}

Using values of $s_{na}$, $s_{ps,n}$ and $s_{a,n}$, we calculate
$\bar{s}_{na}$ and $\bar{s}_{a,n}$ from (\ref{eq:SFG1}) and (\ref{eq:SFG2}).
PU packet throughput for ARC method $\mu_{P,ARC}$ is obtained by
substituting values of $\bar{s}_{na}$, $s_{ps,n}$ and $s_{a,n}$
in (\ref{eq:thru1}).

\subsubsection{R-BRC}

R-BRC differs from ARC only in the retransmission phase. Thus, probability
of successful transmission without assistance $s_{na}$ and probability
of $n$ SUs receiving PU packet $s_{ps,n}$ remains same as derived
in (\ref{eq:Sna}) and (\ref{eq:Spsn}) for ARC. In R-BRC, only the
best SU relay transmits in retransmission phase. Then similar to (\ref{eq:San}),
we can write $s_{a,n}$ as 
\begin{align}
s_{a,n} & =\sum_{m=0}^{L-n}\binom{L-n}{m}\,\left[\left(1-p_{d}\right)q\right]^{m}\nonumber \\
 & \qquad\times\left[1-\left(1-p_{d}\right)q\right]^{L-n-m}\mathcal{U}_{b}\left(n,m\right).\label{eq:San_BRC}
\end{align}
Using (\ref{eq:San_BRC}), PU throughput in R-BRC $\mu_{P,R-BRC}$
is calculated from (\ref{eq:thru1}).

\subsubsection{NR-BRC}

In NR-BRC, unlike previous two methods, assisting SUs discard PU packet
after best SU relay selection. If assisted retransmission is unsuccessful,
in fresh attempt state, the same SU assists while all other SUs continue
sensing spectrum and may interfere. Here values of $\bar{s}_{na}$,
$s_{ps,n}$ and $s_{a,n}$ remain unchanged from the case of R-BRC.
Overall probability that PU transmission is successful when one SU
assists in fresh attempt state is given by 
\begin{align}
s_{f} & =\sum_{m=0}^{L-1}\binom{L-1}{m}\,\left[\left(1-p_{d}\right)q\right]^{m}\nonumber \\
 & \qquad\times\left[1-\left(1-p_{d}\right)q\right]^{L-1-m}\mathcal{U}_{b}\left(1,m\right).\label{eq:Sf}
\end{align}
Using (\ref{eq:Sf}) we get PU throughput in NR-BRC $\mu_{P,NR-BRC}$
from (\ref{eq:thru2}).

\subsubsection{No cooperation (NC)}

As a baseline for performance comparison, we consider a system with
no cooperation between SUs and PU. In this case, SUs do not receive
packets from PU. Thus, there is no assistance in transmission of unsuccessful
PU packets. In this case, probability of successful packet transmission
is same as $s_{na}$ from (\ref{eq:Sna}) and is given by 
\begin{equation}
\mu_{P,NC}=\exp\left(\frac{-\sigma_{N}^{2}\beta}{P_{P}\sigma_{PD}^{2}}\right)\left[1-\frac{B_{P}\left(1-p_{d}\right)q}{1+B_{P}}\right]^{L}.
\end{equation}

It can be seen that PU packet departure process depends on three events--
direct transmission of PU packet without assistance, transmission
from PU to SUs and assisted transmission to PU destination. These
events in turn depend on channel fading processes and events of misdetection
which are stationary. Consequently, PU packet departure process is
stationary.

\subsection{Lower bound on secondary user throughput\label{sub:SUthru}}

For given packet arrival rate $\lambda_{P}$ and packet throughput
$\mu_{P}$, PU queue is stable if $\lambda_{P}<\mu_{P}$. A stable
queue is non-empty with probability $\lambda_{P}/\mu_{P}$ \cite{Kleinrock}.
Thus, PU is silent with probability $\left(1-\lambda_{P}/\mu_{P}\right)$.
SU packet throughput achieved in absence of interference from PU is
of interest as explained in Section II-D.

When PU is inactive, each SU transmits with probability $q$ when
there is no false alarm. Then packet throughput of a single SU is
\begin{align*}
\mu_{S} & =\left(1-\frac{\lambda_{P}}{\mu_{P}}\right)\sum_{m=0}^{L-1}\binom{L-1}{m}\left[\left(1-p_{f}\right)q\right]^{m}\\
 & \qquad\times\left[1-\left(1-p_{f}\right)q\right]^{L-1-m}\mathcal{V}\left(m\right).
\end{align*}
Using expression for $\mathcal{V}\left(m\right)$ as derived in \cite[eq.(17)]{fanous}
and solving, we get 
\begin{equation}
\mu_{S}\!=\!\left(\!1-\frac{\lambda_{P}}{\mu_{P}}\!\right)\!q\!\left(1-p_{f}\!\right)\exp\!\left(\frac{-\sigma_{N}^{2}\beta}{P_{S}\sigma_{SR}^{2}}\right)\!\!\left[1-\frac{q\left(1-p_{f}\right)\beta}{1+\beta}\right]^{L-1}.\label{eq:MuS}
\end{equation}
We summarize the process of calculating PU and SU packet throughput
in IS case in Algorithm $1$.

\begin{algorithm}
\begin{enumerate}
\item Calculate values of branch gains $s_{na}$, $s_{ps,n}$, $s_{a,n}$
and $s_{f}$ using Eq. (\ref{eq:Sna}), (\ref{eq:Spsn}), (\ref{eq:San})
for ARC, Eq. (\ref{eq:Sna}), (\ref{eq:Spsn}), (\ref{eq:San_BRC})
for R-BRC and Eq. (\ref{eq:Sna}), (\ref{eq:Spsn}), (\ref{eq:San_BRC}),
(\ref{eq:Sf}) for NR-BRC. 
\item Calculate PU throughput for ARC/R-BRC using (\ref{eq:thru1}) and
for NR-BRC using (\ref{eq:thru2}). 
\item Calculate SU throughput using (\ref{eq:MuS}).
\end{enumerate}
\caption{Calculating PU and SU packet throughput in IS case\label{alg:algo}}
\end{algorithm}

To keep the system stable, transmission probability $q$ should be
chosen optimally such that SU packet throughput in (\ref{eq:MuS})
is maximized while ensuring PU queue stability. Due to various non-linearities
involved, finding closed form expression of optimal transmission probability
$q^{*}$ is complicated. However $q^{*}$ can be found numerically
as explained later in Section \ref{sub:Effect-on-SU}. For the special
case where $p_{d}=1$, $\mu_{P}$ is independent of $q$. Thus, by
differentiating (\ref{eq:MuS}) and equating to zero, we get 
\begin{equation}
q^{*}=\min\left[\frac{1+\beta}{\left(1-p_{f}\right)\beta L},\,1\right].
\end{equation}

\section{Throughput analysis - Cooperative sensing (CS)}

In cooperative sensing case, SUs share their sensing data with one
of the SUs that acts as fusion center. In this case, there is no interference
in assisted transmission as cooperating SUs can direct non-cooperating
SUs to stay silent. With this property of the CS case, we analyze
PU and SU throughput using results in (\ref{eq:thru1}) and (\ref{eq:thru2}).
It is worth noting that method of cooperative sensing (hard data fusion
or soft data fusion) does not change the ensuing analysis as the analysis
only uses detection probability $p_{d}^{*}$ and false alarm probability
$p_{f}^{*}$. Values of $p_{d}^{*}$ and $p_{f}^{*}$ may change depending
on the underlying CS technique \cite{Akyildiz}.

\subsection{Primary user throughput}

\subsubsection{ARC}

If SUs sense PU presence correctly, there is no interference to PU
transmission. In case of misdetection, each SU transmits and interferes
with probability $q$. Thus, probability of successful PU packet transmission
without assistance is 
\begin{align}
s_{na} & =p_{d}^{*}\,\mathcal{U}_{a}\left(0,0\right)\nonumber \\
 & \qquad+\left(1-p_{d}^{*}\right)\sum_{m=0}^{L}\binom{L}{m}\,q^{m}\left(1-q\right)^{L-m}\mathcal{U}_{a}\left(0,m\right).\label{eq:Sna_CS_pre}
\end{align}
Substituting value of $\mathcal{U}_{a}\left(0,m\right)$ from (\ref{eq:Qd_no_assist}),
we get 
\begin{equation}
s_{na}=\exp\left(\frac{-\sigma_{N}^{2}\beta}{P_{P}\sigma_{PD}^{2}}\right)\left[p_{d}^{*}+\left(1-p_{d}^{*}\right)\left(1-\frac{B_{P}q}{1+B_{P}}\right)^{L}\right],\label{eq:Sna_CS}
\end{equation}
where $B_{P}=\frac{\beta P_{S}\sigma_{SD}^{2}}{P_{P}\sigma_{PD}^{2}}$.

If PU is correctly detected, all SUs try to receive PU packet. Thus,
probability that $n$ SUs receive PU packet when direct transmission
of PU is unsuccessful is 
\begin{equation}
s_{ps,n}=p_{d}^{*}\left(1-\mathcal{U}_{a}\left(0,0\right)\right)\binom{L}{n}\left[\mathcal{W}\left(0\right)\right]^{n}\left[1-\mathcal{W}\left(0\right)\right]^{L-n}.\label{eq:Spsn_CS}
\end{equation}

As there is no interference in cooperation phase, probability of successful
PU packet transmission when $n$ SUs assist is 
\begin{equation}
s_{a,n}=\mathcal{U}_{a}\left(n,0\right).\label{eq:San_CS}
\end{equation}
Using (\ref{eq:SFG1}), (\ref{eq:Sna_CS}), (\ref{eq:Spsn_CS}) and
(\ref{eq:San_CS}), we get PU throughput of ARC from (\ref{eq:thru1}).

\subsubsection{R-BRC}

In this case, values of $s_{na}$, $s_{ps,n}$ remain same as in the
case of ARC. When $n$ SUs have received unsuccessful PU packet, only
the best SU cooperates without any interference. Thus, we have 
\begin{equation}
s_{a,n}=\mathcal{U}_{b}\left(n,0\right).\label{eq:San_BRC_CS}
\end{equation}
PU throughput for R-BRC can be found using (\ref{eq:thru1}).

\subsubsection{NR-BRC}

In this case, values of $s_{na}$, $s_{ps,n}$ and $s_{a,n}$ remain
same as in the case of R-BRC. As there is no interference in cooperation
phase, probability of successful transmission in fresh attempt state
is 
\begin{equation}
s_{f}=\mathcal{U}_{b}\left(1,0\right).\label{eq:Sf_CS}
\end{equation}
PU throughput for NR-BRC can be found using (\ref{eq:thru2}).

\subsubsection{No cooperation (NC)}

In this case, PU packet throughput is same as $s_{na}$ from (\ref{eq:Sna_CS})
and is given by 
\begin{equation}
\mu_{P,NC}=\exp\left(\frac{-\sigma_{N}^{2}\beta}{P_{P}\sigma_{PD}^{2}}\right)\left[p_{d}^{*}+\left(1-p_{d}^{*}\right)\left(1-\frac{B_{P}q}{1+B_{P}}\right)^{L}\right].
\end{equation}
Similar to IS case, PU packet departure process in CS case is a stationary
process as it is a function of stationary events.

\subsection{Lower bound on secondary user throughput}

When PU is inactive, all SUs transmit with probability $q$ if there
is no false alarm. Then SU throughput is given by 
\begin{align}
\mu_{S} & =\left(1-\frac{\lambda_{P}}{\mu_{P}}\right)\left(1-p_{f}^{*}\right)\sum_{m=0}^{L-1}\binom{L-1}{m}\nonumber \\
 & \qquad\qquad\times q^{m}\left(1-q\right)^{L-1-m}\mathcal{V}\left(m\right).\label{eq:MuS_CS_pre}
\end{align}
After simplifying (\ref{eq:MuS_CS_pre}), we get 
\begin{equation}
\mu_{S}=\left(1-\frac{\lambda_{P}}{\mu_{P}}\right)q\left(1-p_{f}^{*}\right)\exp\left(\frac{-\sigma_{N}^{2}\beta}{P_{S}\sigma_{SR}^{2}}\right)\left[1-\frac{\beta q}{1+\beta}\right]^{L-1}.\label{eq:MuS_CS}
\end{equation}

For special case where $p_{d}^{*}=1$, optimal $q$ that maximizes
$\mu_{S}$ is independent of false alarm probability $p_{f}^{*}$
and is given by 
\[
q^{*}=\min\left[\frac{1+\beta}{\beta L},\,1\right].
\]
As process of calculating PU and SU packet throughput in CS case is
similar to IS case as given in Algorithm \ref{alg:algo}, we omit
the algorithm for CS case for brevity.

\section{Numerical results and discussion}

We now present numerical results to study performance of proposed
cooperation methods. Parameter values used are as follows. Transmit
powers of PU and SU are $P_{P}=P_{S}=0.1\,W$. Noise power is $\sigma_{N}^{2}=0.1\,W$.
Average channel gains are- $\sigma_{PD}^{2}=-13\,dB$, $\sigma_{PR}^{2}=0\,dB$,
$\sigma_{PS}^{2}=-10\,dB$, $\sigma_{SR}^{2}=-10\,dB$ and $\sigma_{SD}^{2}=-10\,dB$,
unless stated otherwise. Value of SINR threshold for successful packet
transmission is $\beta=0.1$. Detection probability is $p_{d}=0.8$
and false alarm probability is $p_{f}=0.1$ unless mentioned otherwise.
To demonstrate performance of cooperative sensing (CS) case, we use
majority rule for hard decision combining \cite{Peh}. In majority
rule, probability of detection or false alarm is given by 
\[
p_{j}^{*}=\sum_{n=\left\lceil \frac{L}{2}\right\rceil }^{L}\binom{L}{n}\,p_{j}^{n}\,\left(1-p_{j}\right)^{L-n},\,\,j\in\left\{ d,\,f\right\} .
\]

\subsection{PU throughput}

\subsubsection{Effect of cooperation}

Fig. \ref{fig:MuPvsL} plots PU throughput versus number of SUs for
different cooperation methods. It shows that as number of SUs increases,
more SUs are available for cooperation, resulting in increased PU
packet throughput. ARC performs the best compared to R-BRC and NR-BRC.
This is because all SUs participate in retransmission in case of ARC.
NR-BRC performs worse than R-BRC. This is because, in NR-BRC the same
SU participates in subsequent retransmissions of the same packet,
even if it is not the best SU relay. However, all three methods result
in higher PU throughput than no cooperation (NC) case.

\subsubsection{Effect of sensing}

\begin{figure}
\centering

\includegraphics[scale=0.45]{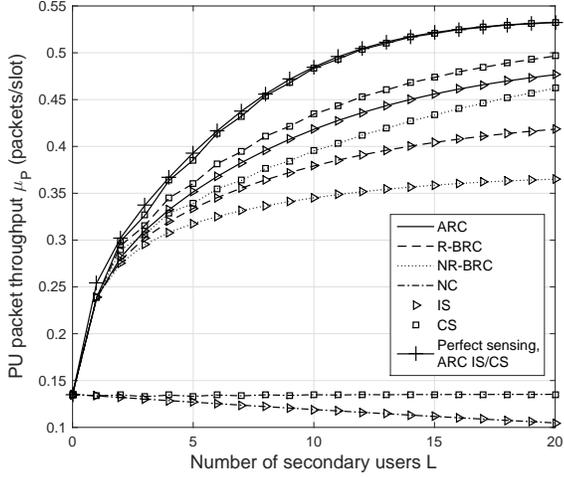}

\protect\protect\protect\caption{PU packet throughput $\mu_{P}$ versus number of SUs $L$ for $q=0.7$
and $\sigma_{SD}^{2}=-13dB$ \label{fig:MuPvsL}}
\end{figure}

Fig. \ref{fig:MuPvsL} also shows that PU throughput in perfect sensing
case is better than imperfect sensing case. This is because, there
is no interference to PU transmission or assisted retransmissions
in perfect sensing case. When SUs employ cooperative sensing, probability
of detection is higher than that in IS case. Thus, the performance
of ARC CS is close to the perfect sensing case.

\subsubsection{Effect of SU transmit power}

\begin{figure}
\centering

\includegraphics[scale=0.45]{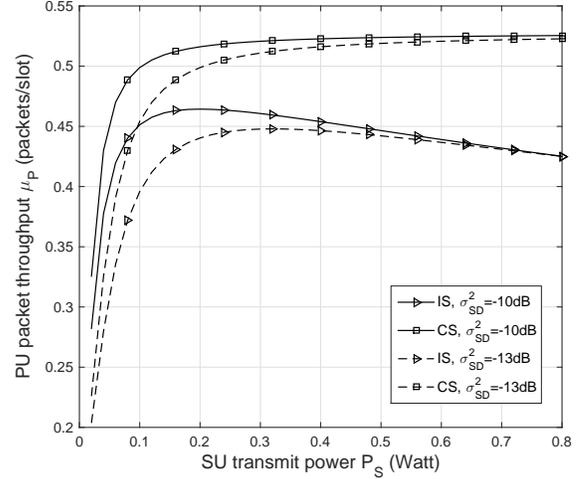}

\protect\protect\protect\caption{Effect of SU transmit power $P_{S}$ on PU packet throughput $\mu_{P}$
for $L=8$ \label{fig:MuPvsPs_compare}}
\end{figure}

When cooperating SUs transmit with higher power, received SINR at
PU destination increases, resulting in higher PU throughput. But interference
caused to PU transmission and assisted retransmissions also increases
as misdetecting SUs transmit with high power. This tradeoff is shown
in Fig. \ref{fig:MuPvsPs_compare}. For IS case, PU throughput initially
increases with increasing $P_{S}$. As $P_{S}$ increases further,
effect of increased interference dominates effect of cooperation and
$\mu_{P}$ decreases. In CS case, effect of interference dominates
effect of cooperation only at very high values of $P_{S}$. When channel
gains between SU sources to PU destination are high, $\mu_{P}$ increases
due to better cooperation. But the tradeoff point is reached sooner
and rate of decrease in $\mu_{P}$ is higher.

\subsection{SU throughput\label{sub:Effect-on-SU}}

\subsubsection{Optimal SU transmission probability}

\begin{figure}
\centering

\includegraphics[scale=0.45]{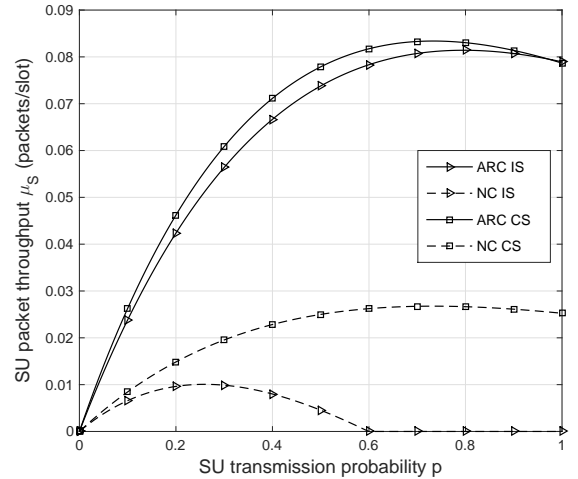}

\protect\protect\protect\caption{Effect of SU transmission probability $q$ on SU packet throughput
$\hat{\mu}_{S}$ for $\lambda_{P}=0.1$ and $L=15$\label{fig:MuSvsP}}
\end{figure}

In case of imperfect sensing, PU packet throughput $\mu_{P}$ is a
function of $q$. For packet arrival rate $\lambda_{P}$, system becomes
unstable if $\mu_{P}\left(q\right)<\lambda_{P}$. To find optimal
$q$ that maximizes SU packet throughput and ensures PU queue stability,
we define SU packet throughput by an auxiliary function as 
\[
\hat{\mu}_{S}=\begin{cases}
\mu_{S} & \mbox{for}\,\,\,\mu_{P}\left(q\right)\geq\lambda_{P}\\
0 & \mbox{for}\,\,\,\mu_{P}\left(q\right)<\lambda_{P}
\end{cases},
\]
where $\mu_{S}$ is the lower bound on SU throughput derived in (\ref{eq:MuS})
for IS case and in (\ref{eq:MuS_CS}) for CS case. It can be seen
that $q$ that maximizes $\hat{\mu}_{S}$ also maximizes $\mu_{S}$
and ensures queue stability of PU. Fig. \ref{fig:MuSvsP} plots $\hat{\mu}_{S}$
versus transmission probability $q$ for $\lambda_{P}=0.1$ and $L=15$.
With increasing $q$, SU transmits more often and achieves higher
packet throughput. But higher value of $q$ results in decreased PU
throughput which lowers probability of PU being silent. Thus, $\hat{\mu}_{S}$
falls when $q$ increases further. This tradeoff is seen for cooperation
as well as non-cooperation cases. When $\mu_{P}\left(q\right)$ falls
below $\lambda_{P}$, system becomes unstable and $\hat{\mu}_{S}$
becomes zero. Using this tradeoff, optimal $q$ that maximizes $\hat{\mu}_{S}$
is found by numerical search.

\subsubsection{Effect of cooperation}

Fig. \ref{fig:MuSvsL} plots lower bound on SU packet throughput derived
in (\ref{eq:MuS}) and (\ref{eq:MuS_CS}) versus number of SUs $L$
for IS and CS case. With increase in number of SUs, PU throughput
increases, resulting in more silent slots. This results in improvement
in SU throughput initially. However as $L$ increases further, inter-SU
interference becomes dominant and $\mu_{S}$ decreases. Similar to
the PU throughput performance ARC performs better than R-BRC and NR-BRC.
NR-BRC performs worse than R-BRC.

\subsubsection{Effect of sensing}

\begin{figure}
\centering

\includegraphics[scale=0.45]{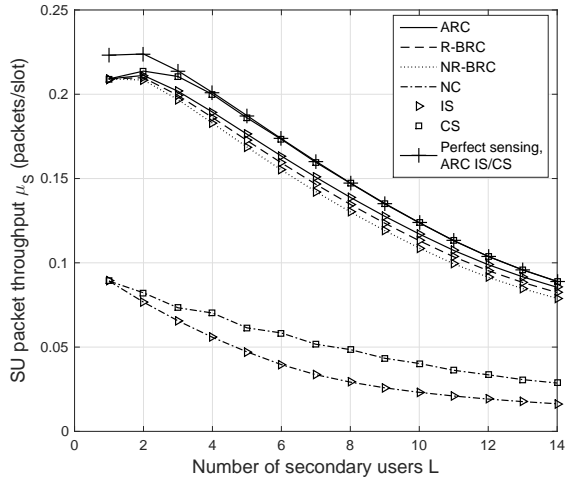}

\protect\protect\protect\caption{SU packet throughput $\mu_{S}$ versus number of SUs $L$ for $p_{f}=0.02$,
$\lambda_{P}=0.1$ and $\sigma_{SD}^{2}=-13dB$\label{fig:MuSvsL}}
\end{figure}

In case of imperfect sensing, PU throughput is less than that in perfect
sensing case. This results in lower probability of PU queue being
empty. Also, with non-zero $p_{f}$, SUs sense some silent slots as
being active and do not transmit. Thus, SU throughput for imperfect
sensing case is less as shown in Fig. \ref{fig:MuSvsL}. In cooperative
sensing, $p_{d}^{*}$ is high and $p_{f}^{*}$ is significantly low.
Thus, SU throughput in CS case is close to the perfect sensing case.

\subsection{Stable throughput region}

\begin{figure}
\centering

\includegraphics[scale=0.45]{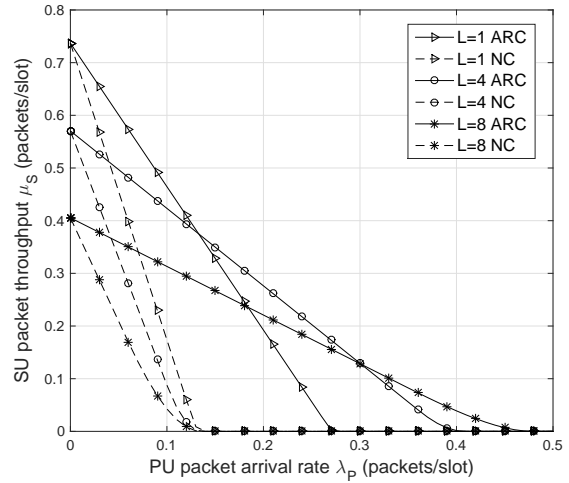}

\protect\protect\protect\caption{SU packet throughput $\mu_{S}$ vs PU packet arrival rate $\lambda_{P}$
for individual sensing case for $\sigma_{SR}^{2}=-7\,dB$\label{fig:MuSvsLem}}
\end{figure}

All tuples of PU packet arrival rate $\lambda_{P}$ and SU packet
throughput $\mu_{S}$ that keep the system stable make up the stable
throughput region. Fig. \ref{fig:MuSvsLem} plots SU packet throughput
$\mu_{S}$ versus PU packet arrival rate $\lambda_{P}$ for IS case.
With large number of SUs, PU throughput increases. Thus, PU can support
higher packet arrival rate $\lambda_{P}$ while keeping the system
stable. However, as inter-SU interference increases with increase
in number of SUs, maximum achievable $\mu_{S}$ (at $\lambda_{P}=0$)
decreases. In contrast, for NC case, maximum SU throughput $\mu_{S}$
as well as maximum supported $\lambda_{P}$ decrease with increasing
$L$.

\subsection{Average PU packet delay: Comparison with \cite{Ashour}}

\begin{figure}
\centering

\includegraphics[scale=0.45]{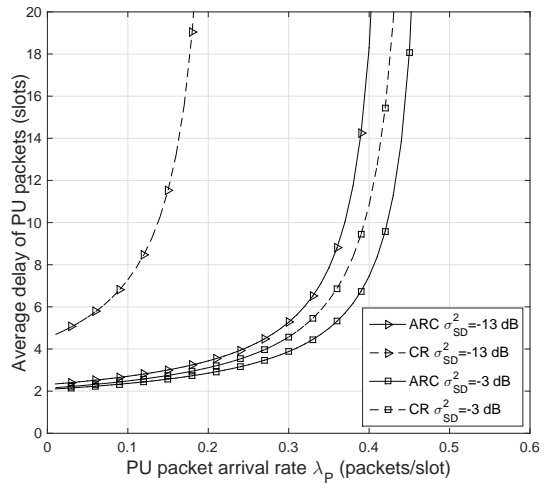}

\protect\protect\protect\caption{Comparison of average PU packet delay in ARC and cooperative relaying
(CR) protocol in \cite{Ashour} for perfect sensing case\label{fig:MuSvsLem_Fanous}}
\end{figure}

In Fig. \ref{fig:MuSvsLem_Fanous}, we compare average delay performance
of ARC with cooperative relaying (CR) protocol in \cite{Ashour} under
perfect sensing case for $L=1$. Average PU packet delay in CR has
been derived in \cite[Eq. (14), (16), (17)]{Ashour}. Similarly, using
Little's law and Pollaczek-Khinchine formula \cite{Kleinrock}, average
delay experienced by PU packets in proposed cooperation method can
be written as 
\[
D_{P}=\frac{1-\lambda_{P}}{\mu_{P}-\lambda_{P}}.
\]

In CR, some PU packets get queued up in two queues-- PU queue and
relay queue at SU-- before reaching PU destination. Also SU relays
PU packets only when PU queue is empty. This results in greater delay
as compared to ARC. When SU source to PU destination link is weak,
improvement in delay performance due to ARC is significant.

\section{Conclusion}

In this paper, we proposed a cooperation method where unsuccessful
PU packets are retransmitted by PU as well as cooperating SUs. Depending
on how SUs are chosen for cooperation, three variations of the cooperation
method were presented. We analyzed packet throughput of PU and SU
in these methods by representing them as signal flow graph and using
graph reduction tools. Individual sensing as well as cooperative sensing
cases were considered. The cooperation methods result in significant
packet throughput gains over systems with no cooperation. These performance
gains are achieved by using minimum resources for relay queue at each
SU. It was observed that higher number of SUs offer more cooperation
to PU and result in higher PU packet throughput. But individual SU
throughput decreases due to increased inter-SU interference. Optimal
transmission probability to maximize SU throughput and keep PU queue
stable can be found numerically.

\appendices{}

\section{Derivation of success probability $\mathcal{U}_{a}\left(n,m\right)$\label{sec:Prob_ARC}}

\numberwithin{equation}{section} \setcounter{equation}{0}

In ARC, all SUs that have received unsuccessful PU packet assist in
PU transmission using D-OSTBC. When $n>0$ SUs assist and $m>0$ SUs
interfere, we have 
\begin{equation}
\mathcal{U}_{a}\left(n,m\right)=\Pr\left[\frac{P_{P}\left|h_{PD}\right|^{2}+\sum_{i=1}^{n}P_{S}\left|h_{S_{i}D}\right|^{2}}{\sigma_{N}^{2}+\sum_{j=1}^{m}P_{S}\left|h_{S_{j}D}\right|^{2}}>\beta\right].\label{eq:Qd}
\end{equation}
We write (\ref{eq:Qd}) in the form 
\begin{equation}
\mathcal{U}_{a}\left(n,m\right)=\Pr\left[W>c+X-Y\right],\label{eq:Qd_rv}
\end{equation}
where 
\begin{itemize}
\item $X=\beta\sum_{j=1}^{m}P_{S}\left|h_{S_{j}D}\right|^{2}$ is a Gamma
random variable with probability density function (PDF) $f_{X}\left(x\right)=\frac{\beta_{1}^{m}}{\Gamma\left(m\right)}x^{m-1}e^{-\beta_{1}x}$
with shape parameter $m$ and rate parameter $\beta_{1}=\frac{1}{\beta_{P}P_{S}\sigma_{SD}^{2}}$
where $\Gamma\left(\cdot\right)$ is Gamma function defined as $\Gamma\left(a\right)=\int_{0}^{\infty}x^{a-1}e^{-x}dx$
\cite{papoulis2002probability}, 
\item $Y=\sum_{i=1}^{n}P_{S}\left|h_{S_{i}D}\right|^{2}$ is a Gamma random
variable with PDF $f_{Y}\left(y\right)=\frac{\beta_{2}^{n}}{\Gamma\left(n\right)}y^{n-1}e^{-\beta_{2}y}$
with shape parameter $n$ and rate parameter $\beta_{2}=\frac{1}{P_{S}\sigma_{SD}^{2}}$, 
\item $W=P_{P}\left|h_{PD}\right|^{2}$ is an exponential random variable
with PDF $f_{W}\left(w\right)=\beta_{3}e^{-\beta_{3}w}$ with rate
parameter $\beta_{3}=\frac{1}{P_{P}\sigma_{PD}^{2}}$, 
\item $c=\beta\sigma_{N}^{2}$ is a constant. 
\end{itemize}
To calculate probability in (\ref{eq:Qd_rv}), we first find distribution
of difference of Gamma distributed independent random variables $X$
and $Y$. Let $Z=X-Y$. As $X,\,Y\in\left[0,\,\infty\right)$, $Z=X-Y$
takes on values in $\left(-\infty,\,\infty\right)$. We have 
\begin{equation}
f_{X+Y}\left(z\right)=\int_{-\infty}^{\infty}f_{X}\left(x\right)f_{Y}\left(z-x\right)dx.\label{eq:Convolution}
\end{equation}
Using (\ref{eq:Convolution}) and noting that $f_{-Y}\left(y\right)=f_{Y}\left(-y\right)$,
we get 
\begin{align}
f_{X-Y}\left(z\right) & =\int_{-\infty}^{\infty}f_{X}\left(x\right)f_{-Y}\left(z-x\right)dx\nonumber \\
 & =\int_{-\infty}^{\infty}f_{X}\left(x\right)f_{Y}\left(x-z\right)dx.\label{eq:Convolution2}
\end{align}
We can further write (\ref{eq:Convolution2}) as 
\begin{equation}
f_{Z}\left(z\right)=\begin{cases}
\int_{0}^{\infty}f_{X}\left(x\right)f_{Y}\left(x-z\right)dx & \,\,\mbox{for}\,z<0\\
\int_{0}^{\infty}f_{X}\left(y+z\right)f_{Y}\left(y\right)dy & \,\,\mbox{for}\,z\geq0
\end{cases}.\label{eq:Convolution3}
\end{equation}
Substituting expressions of PDFs of $X$ and $Y$ in (\ref{eq:Convolution3})
and simplifying we get PDF of $Z$ as given in (\ref{eq:PDF_final})
on next page.

In the original problem of (\ref{eq:Qd_rv}), $W$ is exponentially
distributed, hence $W\in\left[0,\,\infty\right)$. Thus, we can write
\[
\Pr\left[W>c+Z\right]=\begin{cases}
1 & \,\,\mbox{for}\,Z<-c\\
e^{-\beta_{3}\left(c+Z\right)} & \,\,\mbox{for}\,Z\geq-c
\end{cases}.
\]
Thus, overall probability of successful transmission in (\ref{eq:Qd})
is given as 
\begin{align}
\mathcal{U}_{a}\left(n,m\right) & =\underbrace{\int_{-\infty}^{-c}f_{Z}\left(z\right)dz}_{I_{A_{1}}}+\underbrace{\int_{-c}^{0}e^{-\beta_{3}\left(c+z\right)}f_{Z}\left(z\right)dz}_{I_{A_{2}}}\nonumber \\
 & \qquad+\underbrace{\int_{0}^{\infty}e^{-\beta_{3}\left(c+z\right)}f_{Z}\left(z\right)dz}_{I_{A_{3}}},\label{eq:Qd_final}
\end{align}
where $I_{A_{1}}$, $I_{A_{2}}$ and $I_{A_{3}}$ are found using
(\ref{eq:PDF_final}). Let $\Gamma\left(a,\,x\right)$ be the upper
incomplete Gamma function defined as $\Gamma\left(a,\,x\right)=\int_{x}^{\infty}t^{a-1}e^{-t}dt$.
Also let $\gamma\left(a,\,x\right)$ be the lower incomplete Gamma
function defined as $\gamma\left(a,\,x\right)=\int_{0}^{x}t^{a-1}e^{-t}dt$.
Then the expressions for $I_{A_{1}}$, $I_{A_{2}}$, $I_{A_{3}}$
are as given in (\ref{eq:I1}), (\ref{eq:I2}) and (\ref{eq:I3})
on next page.

When no SU interferes $\left(m=0\right)$ and $n>0$ SUs assist, using
similar approach, we get probability of successful transmission as
\begin{equation}
\mathcal{U}_{a}\left(n,0\right)\!=\!\begin{cases}
\frac{\Gamma\left(n,\,\beta_{2}c\right)}{\Gamma\left(n\right)}+e^{-\beta_{3}c}\frac{\beta_{2}^{n}}{\Gamma\left(n\right)}\frac{\gamma\left(n,\,\left(\beta_{2}-\beta_{3}\right)c\right)}{\left(\beta_{2}-\beta_{3}\right)^{n}}\\
\qquad\qquad\qquad\qquad\qquad\qquad\mbox{for}\,\beta_{2}\neq\beta_{3}\\
\frac{\Gamma\left(n,\,\beta_{2}c\right)}{\Gamma\left(n\right)}+e^{-\beta_{3}c}\frac{\beta_{2}^{n}}{\Gamma\left(n\right)}\frac{c^{n}}{n}\\
\qquad\qquad\qquad\qquad\qquad\qquad\mbox{for}\,\beta_{2}=\beta_{3}
\end{cases}.\label{eq:Qd_no_intf}
\end{equation}

When no SU assists $\left(n=0\right)$ but $m\geq0$ SUs interfere,
probability of successful transmission is same as that derived in
\cite[Eq.(11)]{fanous} and is given by 
\begin{align}
\mathcal{U}_{a}\left(0,m\right) & =e^{-\beta_{3}c}\left[1+\frac{\beta_{3}}{\beta_{1}}\right]^{-m}\nonumber \\
 & =\exp\left(\frac{-\sigma^{2}\beta}{P_{P}\sigma_{PD}^{2}}\right)\left[1+\frac{\beta P_{S}\sigma_{SD}^{2}}{P_{P}\sigma_{PD}^{2}}\right]^{-m}.\label{eq:Qd_no_assist}
\end{align}

\begin{figure*}
\begin{equation}
f_{Z}\left(z;\,m,\,\beta_{1},\,n,\,\beta_{2}\right)=\begin{cases}
\frac{\beta_{1}^{m}\beta_{2}^{n}}{\Gamma\left(m\right)\Gamma\left(n\right)}e^{-\beta_{1}z}\sum_{j=0}^{m-1}\binom{m-1}{j}\,z^{j}\frac{\Gamma\left(m+n-1-j\right)}{\left(\beta_{1}+\beta_{2}\right)^{m+n-1-j}} & \,\,\mbox{for}\,z\geq0\\
\frac{\beta_{1}^{m}\beta_{2}^{n}}{\Gamma\left(m\right)\Gamma\left(n\right)}e^{\beta_{2}z}\sum_{j=0}^{n-1}\binom{n-1}{j}\,\left|z\right|^{j}\frac{\Gamma\left(m+n-1-j\right)}{\left(\beta_{1}+\beta_{2}\right)^{m+n-1-j}} & \,\,\mbox{for}\,z<0
\end{cases}\label{eq:PDF_final}
\end{equation}

\rule[0.5ex]{2\columnwidth}{0.5pt} 
\end{figure*}

\begin{figure*}
\begin{equation}
I_{A_{1}}=\frac{\beta_{1}^{m}\beta_{2}^{n}}{\Gamma\left(m\right)\Gamma\left(n\right)}\sum_{j=0}^{n-1}\binom{n-1}{j}\,\frac{\Gamma\left(m+n-1-j\right)}{\left(\beta_{1}+\beta_{2}\right)^{m+n-1-j}}\frac{\Gamma\left(j+1,\,\beta_{2}c\right)}{\beta_{2}^{j+1}}\label{eq:I1}
\end{equation}
\begin{equation}
I_{A_{2}}=\begin{cases}
\frac{\beta_{1}^{m}\beta_{2}^{n}}{\Gamma\left(m\right)\Gamma\left(n\right)}e^{-\beta_{3}c}\sum_{j=0}^{n-1}\binom{n-1}{j}\,\frac{\Gamma\left(m+n-1-j\right)}{\left(\beta_{1}+\beta_{2}\right)^{m+n-1-j}}\frac{\gamma\left(j+1,\,\left(\beta_{2}-\beta_{3}\right)c\right)}{\left(\beta_{2}-\beta_{3}\right)^{j+1}} & \,\,\mbox{for}\,\beta_{2}\neq\beta_{3}\\
\frac{\beta_{1}^{m}\beta_{2}^{n}}{\Gamma\left(m\right)\Gamma\left(n\right)}e^{-\beta_{3}c}\sum_{j=0}^{n-1}\binom{n-1}{j}\,\frac{\Gamma\left(m+n-1-j\right)}{\left(\beta_{1}+\beta_{2}\right)^{m+n-1-j}}\frac{c^{j+1}}{\left(j+1\right)} & \,\,\mbox{for}\,\beta_{2}=\beta_{3}
\end{cases}\label{eq:I2}
\end{equation}
\begin{equation}
I_{A_{3}}=\frac{\beta_{1}^{m}\beta_{2}^{n}}{\Gamma\left(m\right)\Gamma\left(n\right)}e^{-\beta_{3}c}\sum_{j=0}^{m-1}\binom{m-1}{j}\,\frac{\Gamma\left(m+n-1-j\right)}{\left(\beta_{1}+\beta_{2}\right)^{m+n-1-j}}\frac{\Gamma\left(j+1\right)}{\left(\beta_{1}+\beta_{3}\right)^{j+1}}\label{eq:I3}
\end{equation}

\rule[0.5ex]{2\columnwidth}{0.5pt} 
\end{figure*}

\section{Derivation of success probability $\mathcal{U}_{b}\left(n,m\right)$\label{sec:Prob_BRC}}

\begin{figure*}
\begin{equation}
I_{B_{1}}=\begin{cases}
\beta_{1}^{m}\beta_{2}\times\frac{1-e^{-\left(\beta_{2}-j\beta_{3}\right)c}}{\left(\beta_{2}-j\beta_{3}\right)} & \,\,\mbox{for}\,\beta_{2}\neq j\beta_{3}\\
\beta_{1}^{m}\beta_{2}c & \,\,\mbox{for}\,\beta_{2}=j\beta_{3}
\end{cases}\label{eq:I1_BRC}
\end{equation}
\begin{equation}
I_{B_{2}}=\frac{\beta_{1}^{m}\beta_{2}}{\Gamma\left(m\right)}\sum_{i=0}^{m-1}\binom{m-1}{i}\,\frac{\Gamma\left(m-i\right)}{\left(\beta_{1}+\beta_{2}\right)^{m-i}}\frac{\Gamma\left(i+1\right)}{\left(\beta_{2}+j\beta_{3}\right)^{i+1}}\label{eq:I2_BRC}
\end{equation}

\rule[0.5ex]{2\columnwidth}{0.5pt} 
\end{figure*}

In R-BRC and NR-BRC, out of all SUs that have received PU packet,
the SU having best SU source to PU destination channel is selected
for cooperation. Only the best SU and PU transmit in retransmission
phase using D-OSTBC. Non-assisting SUs may interfere in case of misdetection.
Then probability of successful transmission of PU packet when $n>0$
SUs participate in best relay selection and $m>0$ SUs interfere is
\begin{equation}
\mathcal{U}_{b}\left(n,m\right)=\Pr\left[\frac{P_{P}\left|h_{PD}\right|^{2}+P_{S}\left|h_{SD}^{*}\right|^{2}}{\sigma_{N}^{2}+\sum_{j=1}^{m}P_{S}\left|h_{S_{j}D}\right|^{2}}>\beta\right],\label{eq:Qb}
\end{equation}
where $\left|h_{SD}^{*}\right|^{2}=\max\left\{ \left|h_{S_{1}D}\right|^{2},\,\left|h_{S_{2}D}\right|^{2},\dots,\,\left|h_{S_{n}D}\right|^{2}\right\} $.
Note that in (\ref{eq:Qb}), $\left\{ S_{1},\,S_{2},\dots,\,S_{n}\right\} $
is a set of arbitrary SU sources that have received PU packet and
the subscripts represent no particular order. We write (\ref{eq:Qb})
as 
\begin{equation}
\mathcal{U}_{b}\left(n,m\right)=\Pr\left[W>c+Z\right],\label{eq:Qb_pre}
\end{equation}
where 
\begin{itemize}
\item $W=\max\left\{ X_{1},\,X_{2},\dots,\,X_{n}\right\} $ where $X_{i},\,i=1,\,2,\dots,\,n$
is exponential random variable with PDF $f_{X}\left(x\right)=\beta_{3}e^{-\beta_{3}x}$
with rate parameter $\beta_{3}=\frac{1}{P_{S}\sigma_{SD}^{2}}$, 
\item $Z=\beta\sum_{j=1}^{m}P_{S}\left|h_{S_{j}D}\right|^{2}-P_{P}\left|h_{PD}\right|^{2}$
is difference of two Gamma random variables, with PDF $f_{Z}\left(z;\,m,\,\beta_{1},\,1,\,\beta_{2}\right)$
as given in (\ref{eq:PDF_final}) where $\beta_{1}=\frac{1}{\beta P_{S}\sigma_{SD}^{2}}$
and $\beta_{2}=\frac{1}{P_{P}\sigma_{PD}^{2}}$, 
\item $c=\beta\sigma_{N}^{2}$ is a constant. 
\end{itemize}
Due to independence of $X_{i}$'s, we get cumulative distribution
function (CDF) of $W$ as 
\begin{align}
\Pr\left[W\leq w\right] & =\Pr\left[\max\left\{ X_{1},\,X_{2},\dots,\,X_{n}\right\} \leq w\right]\nonumber \\
 & =\prod_{i=1}^{n}\Pr\left[X_{i}\leq w\right]\nonumber \\
 & =\left(1-e^{\beta_{3}w}\right)^{n}.\label{eq:CDF}
\end{align}
Let $\overline{\mathbb{\mathcal{U}}_{b}}\left(n,m\right)=1-\mathcal{U}_{b}\left(n,m\right)=\Pr\left[W\leq c+Z\right]$.
As $W$ takes values in $\left[0,\,\infty\right)$, we have 
\[
\Pr\left[W\leq c+Z\right]=\begin{cases}
0 & \,\,\mbox{for}\,Z<-c\\
\left(1-e^{-\beta_{3}c}e^{-\beta_{3}Z}\right)^{n} & \,\,\mbox{for}\,Z\geq-c
\end{cases}.
\]
Then overall probability $\overline{\mathbb{\mathcal{U}}_{b}}\left(n,m\right)$
is 
\[
\overline{\mathbb{\mathcal{U}}_{b}}\left(n,m\right)=\int_{-c}^{\infty}\left(1-e^{-\beta_{3}c}e^{-\beta_{3}z}\right)^{n}f_{Z}\left(z\right)\,dz.
\]
Using binomial expansion, we get 
\begin{align}
\overline{\mathbb{\mathcal{U}}_{b}}\left(n,m\right) & =\sum_{j=0}^{n}\binom{n}{j}\left(-1\right)^{j}e^{-j\beta_{3}c}\nonumber \\
 & \!\!\times\!\left[\underbrace{\int_{-c}^{0}e^{-j\beta_{3}z}f_{Z}\left(z\right)\,dz}_{I_{B_{1}}}+\underbrace{\int_{0}^{\infty}e^{-j\beta_{3}z}f_{Z}\left(z\right)\,dz}_{I_{B_{2}}}\right].
\end{align}
Using $f_{Z}\left(z\right)$ from (\ref{eq:PDF_final}), we can write
$I_{B_{1}}$ and $I_{B_{2}}$ as given in (\ref{eq:I1_BRC}) and (\ref{eq:I2_BRC})
on next page.

When no SU interferes $\left(m=0\right)$, using similar approach,
we get 
\begin{equation}
\overline{\mathbb{\mathcal{U}}_{b}}\left(n,0\right)=\sum_{j=0}^{n}\binom{n}{j}\,\left(-1\right)^{j}e^{-j\beta_{3}c}\beta_{2}\mathcal{C}_{j},
\end{equation}
where 
\begin{equation}
\mathcal{C}_{j}=\begin{cases}
\frac{\left[1-e^{-\left(\beta_{2}-j\beta_{3}\right)c}\right]}{\left(\beta_{2}-j\beta_{3}\right)} & \,\,\mbox{for}\,\beta_{2}\neq j\beta_{3}\\
c & \,\,\mbox{for}\,\beta_{2}=j\beta_{3}
\end{cases}.
\end{equation}
Using (\ref{eq:Qb_pre}), we get $\mathcal{U}_{b}\left(n,m\right)=1-\overline{\mathbb{\mathcal{U}}_{b}}\left(n,m\right)$.

 \bibliographystyle{IEEEtran}
\bibliography{database_full,IEEEfull,IEEEabrv}

\end{document}